\documentstyle[aps,prl,twocolumn, epsf]{revtex}

\begin{document}

\draft

\title{Approximate transformations and robust manipulation \\of
bipartite pure state entanglement}

\author{Guifr\'e Vidal$^*$, Daniel Jonathan$^{\dagger}$ and M.~A.~Nielsen$^{\ddagger}$,}

\address{
* Departament d'Estructura i Constituents de la Mat\`eria, Universitat de Barcelona,
Diagonal 647, E-08028 Barcelona, Spain.\\
$\dagger$ Optics Section, the Blackett Laboratory, Imperial College, London SW7 2BZ,
United Kingdom.\\
$\ddagger$ Department of Physics, MC 12-33, California Institute of Technology,
Pasadena, CA 91125, USA.
}

\date{\today}

\maketitle

\begin{abstract}
  We analyze {\em approximate} transformations of pure entangled
  quantum states by local operations and classical communication,
  finding explicit conversion strategies which optimize the fidelity
  of transformation.  These results allow us to determine the most
  faithful teleportation strategy via an initially shared partially
  entangled pure state. They also show that procedures for
  entanglement manipulation such as entanglement catalysis [Jonathan
  and Plenio, Phys. Rev. Lett.  {\bf 83},~3566 (1999)] are robust
  against perturbation of the states involved, and motivate the notion
  of {\em non-local fidelity}, which quantifies the difference in the
  entangled properties of two quantum states.
\end{abstract}


\bigskip

\section{Introduction}

%
%
Entanglement is a {\em resource} at the heart of quantum mechanics;
iron in the classical world's bronze age.  It is a key ingredient in
effects such as quantum computation \cite{Preskill98a}, quantum
teleportation \cite{Bennett93a}, and superdense coding
\cite{Bennett92a}.  To better understand entanglement as a resource,
we would like to understand what transformations of an entangled state
may be accomplished, when only some restricted class of operations is
allowed to accomplish this transformation.  This paradigm, introduced
in \cite{Bennett96b,Bennett96c,Bennett96a}, has been very successful
in identifying many of the fundamental properties of entanglement. The
best studied class of operations is {\em local operations and
classical communication} (LOCC) --- that is, the two entangled parties
may do whatever they wish to their local system, and may communicate
classically, but they cannot use quantum communication.

%
%
This class of transformations has been studied in considerable detail
in~\cite{Lo97b,Nielsen99a,Vidal99a,Jonathan99a,Hardy99a}.  The purpose
of this paper is to generalize earlier results to study {\em
  approximate} transformations of one pure state into another.  In
particular, we obtain a scheme for performing the best possible
entanglement transformation, in the sense that the transformation
results in a state which is ``nearest'' the desired target state, with
respect to a well-motivated measure of distance.  Our results show
that existing results about entanglement transformation are robust
against the effects of slight noise, and quantify exactly how robust.
Our results extend and complement recent and independent work by
Barnum \cite{Barnum99a} on approximate transformations with
applications to cryptography.

%
%
The paper is structured as follows.  In Section~\ref{sec:background}
we review the relevant background material.
Section~\ref{sec:main_result} proves the main result of the paper, an
optimal scheme for performing approximate entanglement transformation.
Section~\ref{sec:discussion} illustrates our main result by
application to some concrete entanglement transformation tasks. In
particular, we determine the optimal fidelity of any teleportation
scheme that uses a partially entangled pure state as its quantum
channel.  Section~\ref{sec:non_local} introduces the concept of {\em
  non-local fidelity} between two entangled states, and studies some
elementary properties of this measure of distance between two
entangled states.  Section~\ref{sec:conc} concludes the paper.

\section{Background}
\label{sec:background}

%
%
Suppose $\psi$ is a pure state of a bipartite system shared by Alice
and Bob, and let
\begin{eqnarray}
|\psi\rangle = \sum_{i=1}^n \sqrt{\alpha_i} |i_A~
i_B\rangle,~~~~\alpha_i\geq \alpha_{i+1}\geq 0,~\sum_{i=1}^n \alpha_i
= 1,
\label{psi}
\end{eqnarray}
be its Schmidt decomposition\cite{Peres93a}.  (Throughout this paper
we switch back and forth between the bra-ket notation $|\psi\rangle$
and the notation $\psi$ without comment.)  Without loss of generality
we may suppose Alice and Bob have state spaces of equal dimension,
$n$.  All results extend trivially to the case of unequal dimensions.
Suppose the parties wish to transform this initial state into a second
pure state $|\phi\rangle$ with Schmidt decomposition,
\begin{eqnarray}
|\phi\rangle = \sum_{i=1}^n \sqrt{\beta_i}
|i_A'~i_B'\rangle,~~~~\beta_i\geq \beta_{i+1}\geq 0,~\sum_{i=1}^n
\beta_i = 1,
\label{phi}
\end{eqnarray}
that we shall call the {\em target} state, by just acting locally on
their subsystems and communicating classically.

Necessary and sufficient conditions for this {\em deterministic} local
transformation to be possible, along with an explicit protocol for the
conversion, were presented in \cite{Nielsen99a}. It was shown there
that $\psi$ is locally convertible into $\phi$ in a deterministic
manner if and only if the vector $\vec{\alpha} = (\alpha_1,\ldots ,
\alpha_n)$ is {\em majorized} by the vector $\vec{\beta} = (\beta_1,
\ldots , \beta_n)$, $\vec{\alpha} \prec \vec{\beta}$:
\begin{eqnarray}
\psi \longrightarrow \phi~~ \Longleftrightarrow ~~\sum_{i=1}^k
\alpha_i \leq \sum_{i=1}^k \beta_i,~~~~k = 1, \cdots, n,
\label{iff}
\end{eqnarray}
with equality holding when $k = n$.  Condition~(\ref{iff}) can be
given an equivalent description in terms of the {\em entanglement
monotones} $E_l$, $l= 1, \ldots, n$, introduced in \cite{Vidal99a},
\begin{eqnarray}
E_l^{\psi} \equiv \sum_{i=l}^n \alpha_i,~~~E_l^{\phi} \equiv \sum_{i=l}^n \beta_i,
\label{eqtn:monotones}
\end{eqnarray}
which are quantities that do not increase, on average, under any local
transformation \cite{Vidal98a}. The state $\psi$ can be locally
transformed into $\phi$ with certainty if and only if none of these
entanglement monotones are increased during the conversion, that is,
\begin{eqnarray}
E_l^{\psi} \geq E_l^{\phi}~~~~l = 1, \cdots, n.
\label{iff2}
\end{eqnarray}

We suppose from now on that condition (\ref{iff}) is not satisfied,
and that therefore the parties cannot locally convert $\psi$ into
$\phi$ deterministically, that is, $\psi \not\!\!\longrightarrow
\phi$. What options do they have?

In some cases, namely when $\psi$ has at least as many non-vanishing
Schmidt coefficients as $\phi$, the parties can still locally
transform $\psi$ into $\phi$ with some non-vanishing probability of
success, performing what we shall call a {\em conclusive}
conversion. The optimal {\em conclusive} protocol is the one with the
maximal probability $P(\psi \rightarrow \phi)$ that the conversion is
successful. This probability can be shown to be\cite{Vidal99a}
\begin{eqnarray}
P(\psi \rightarrow \phi) = \min_{l\in[1,n]} \frac{E_l(\psi)}{E_l(\phi)},
\label{maxprob}
\end{eqnarray}
and thus it is the greatest quantity compatible with the
non-increasing character of the entanglement monotones $E_l$. We can
rephrase this fact by saying that the optimal probability $P(\psi
\rightarrow \phi)$ is the greatest weight $p$ such that $\vec{\alpha}$
is {\em (weakly) submajorized} \cite{Bhatia97a}\footnote{Notice that
whereas in equation~(\ref{iff}) majorization assumes $\sum_{i=1}^n
\alpha_i = \sum_{i=1}^n \beta_i$, which is automatically satisfied due
to the normalization of states $\psi$ and $\phi$, in
equation~(\ref{weak}) the term weak submajorization indicates that the
previous equality has turned into an inequality, $\sum_{i=1}^n
\alpha_i \leq p\sum_{i=1}^n \beta_i$.}
by $p\vec{\beta}$, $\vec{\alpha} \prec_w p\vec{\beta}$, that is,
\begin{eqnarray}
\sum_{i=1}^k \alpha_i \leq p\sum_{i=1}^k \beta_i~~~~k = 1, \cdots, n.
\label{weak}
\end{eqnarray}

An appealing feature of {\em conclusive} conversions is that when
the protocol succeeds the parties end up sharing exactly the target
state $\phi$ they wanted. This is useful in any situation where
Alice and Bob need the target state {\em exactly} and do not wish
to accept a merely {\em similar} outcome, say another state $\xi$
with a reasonably high overlap with $\phi$. One may conceive, for
instance, that the parties want to perform fully reliable
teleportation \cite{Bennett93a}. In order to do so they may try to
conclusively convert the initial pure state $\psi$ into an
$m$-state \cite{Lo97b} --- a state of the form
\begin{eqnarray}
|\psi_m\rangle=\frac{1}{\sqrt{m}}\sum_{i=1}^{m} |i_Ai_B\rangle.
\end{eqnarray}

\begin{figure}
 \epsfysize=3.9cm
\begin{center}
 \epsffile{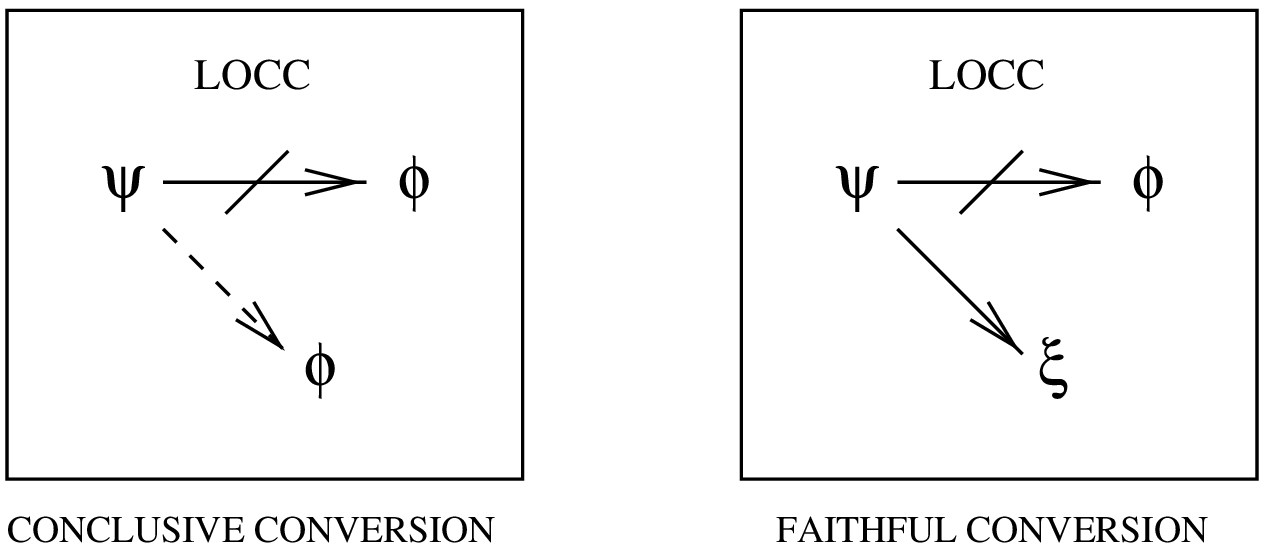}
\end{center}
 \caption{Suppose local operations on the subsystems and classical
 communication between Alice and Bob (LOCC) are not sufficient for a
 {\em deterministic} conversion of the initial state $\psi$ into their
 target state $\phi$, i.e. $\psi \not\rightarrow \phi$. A {\em
 conclusive} local conversion may then do the job with some prior
 probability of success, i.e. sometimes the protocol will lead to the
 target state $\phi$ and sometimes will fail to do so. Alternatively,
 a {\em faithful} conversion will deterministically lead to a final
 state $\xi$ which is only (but often reasonably) similar to the
 target state $\phi$. \label{fig1}}
\end{figure}

In the present work we consider, on the contrary, that the parties
allow for the final outcome $\xi$ of the conversion to be just an
approximated version of the target state $\phi$. We shall call this
alternative type of transformations {\em faithful} (see
Figure~\ref{fig1}). More specifically, we present here faithful
conversions $\psi \longrightarrow \xi$ such that the overlap or
fidelity between the final state $\xi$ and the target state $\phi$,
that is, $F(\xi,\phi) = |\langle \xi| \phi\rangle|^2$, is the greatest
locally achievable.

This {\em approximate} approach is more suitable than conclusive
transformations in a number of contexts. First, it allows us to
consider local conversions when the conclusive ones are not possible
at all. For example, this is the case whenever the target state has
more non-vanishing Schmidt coefficients than the initial
state\cite{Lo97b}. This is relevant to the problem of diluting the
entanglement of a finite set of pure, maximally entangled states into
a larger set of other pure, partially entangled ones.  Such a problem
can be well-posed from the point of view of faithful conversions, and
we will address here the question of which are the optimal (that is,
most faithful) dilution protocols for the finite case.

Entanglement distillation --- the extraction of pure state
entanglement from mixed states --- is a second context where faithful
conversions are highly relevant. It is known \cite{Kent98a} that the
conclusive local conversion of $N$ copies of a mixed state $\rho$ into
any entangled pure state $\phi$ is in general impossible, that is the
probability of making the transformation $\rho^{\otimes N} \rightarrow
\phi$ is typically equal to $0$ for any finite $N$, whereas
distillation of pure state entanglement is often possible in the limit
$N\rightarrow \infty$ \cite{Bennett96a}. Thus faithful conversions of
mixed states into pure states appear as a more adequate framework for
the study of approximate entanglement distillation.

A third reason for interest in faithful transformation protocols is
that, as we will show here, in general they differ from the
conclusive protocols with the highest probability of success.
Finally, the study of approximate conversions allows us to quantify
how robust exact transformations are, a problem of direct relevance
to applications of entanglement transformation such as entanglement
catalysis~\cite{Jonathan99c} and certain cryptographic
protocols~\cite{Barnum99a,Kempe99a}.

\section{Optimal conversions between pure state entanglement}
\label{sec:main_result}

\begin{figure}
 \epsfysize=5.0cm
\begin{center}
 \epsffile{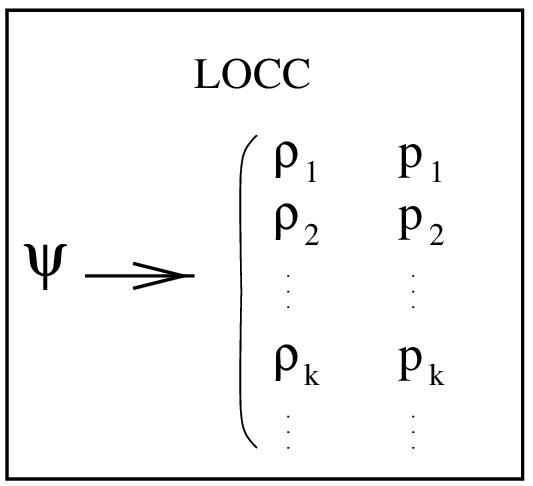}
\end{center}
 \caption{The most general local transformation a bipartite pure state
 $\psi$ can undergo may be {\em probabilistic} in nature, and its
 outcoming states may be {\em mixed}. Lemma~1 allows us to restrict
 our considerations to {\em deterministic} transformations of $\psi$
 into a final {\em pure} state $\xi$, when searching for the most {\em
 faithful} local conversion into a target state $\phi$.  \label{fig2}}
\end{figure}

We consider here the most general local transformations of the initial
state $\psi$, namely those that convert $\psi$ into an ensemble of
possible final states $\rho_k$ with corresponding probabilities $p_k$
(see Figure~(\ref{fig2})). In the case of {\em pure} final states, it
has been shown in \cite{Jonathan99a} that such a {\em probabilistic}
transformation can be performed by local means if, and only if, the
entanglement monotones $E_l$ do not increase on average, that is:
\begin{eqnarray}
\psi \longrightarrow \{p_k, \xi_k\} \Longleftrightarrow E_l^{\psi}
\geq \sum_kp_kE_l^{\xi_k}~~~~l = 1, \cdots, n.
\label{ensemble}
\end{eqnarray}
We can extend this result to the case where the final states may be
mixed states $\rho_k$. Notice that {\em any} local protocol generating
an ensemble $\{p_k, \rho_k\}$ of final mixed states from the pure
state $\psi$ can be (non-uniquely) viewed as the outcome of a two-step
procedure of the following form: first, an ensemble of pure states
$\{p_k q_{k,j}, \xi_{k,j}\}$ such that
\begin{eqnarray}
\rho_k = \sum_j q_{k,j} |\xi_{k,j}\rangle \langle \xi_{k,j}|
\label{mixing}
\end{eqnarray}
is locally produced; then the information concerning the index $j$ is
discarded. Therefore the transformation $\psi \rightarrow \{ p_k,
\rho_k\}$ can be performed locally if, and only if, there exists an
ensemble $\{ p_kq_{k,j}, \xi_{k,j}\}$ satisfying
equation~(\ref{mixing}) and such that
\begin{eqnarray}
 E_l^{\psi} \geq \sum_{k,j}p_{k,j}E_l^{\xi_{k,j}}~~~~l = 1, \cdots, n.
\end{eqnarray}

We can now proceed to the main results of this work. In Lemma~1, we
determine the most faithful strategy for converting between pure
states when only local unitary transformations are allowed. In
Lemma~2, we show that among all possible local transformations of the
initial pure state $\psi$, $\psi \rightarrow\{p_k,\rho_k\}$ (see
Figure~(\ref{fig2})), the maximal average fidelity with respect to the
target state $\phi$, $\sum_k p_k \langle \phi| \rho_k|\phi\rangle$,
can always be obtained in a local and deterministic conversion of the
state $\psi$ into a final pure state $\xi$. These results are then
used to prove Theorem~3, which provides the value of the optimal
fidelity and the identity of the best possible final state $\xi$,
while also constructing an explicit local protocol for the conversion.
It is worth noting that the pure state fidelity is equivalent to the
``trace distance'', a quantity with a well-defined {\em operational
  meaning} as the probability of making an error distinguishing two
states \cite{Fuchs99a}.  The state $\xi$ is in this sense the best
possible physical approximation to the state $\phi$ that may be
achieved using LOCC.  We note that results closely related to Lemma~1
and Lemma~2 have recently been obtained independently by Barnum
\cite{Barnum99a}, however he does not provide the general solution to
the approximation problem, Theorem~3.

%
%

{\bf Lemma 1:} Let $\tau, \omega\in{\cal C}^n\otimes{\cal C}^n$ be two
normalized states with ordered Schmidt decompositions in the same
local basis, that is,
\begin{eqnarray}
|\tau\rangle & = & \sum_{i=1}^n \sqrt{\tau_i} |i_A~i_B\rangle,
~~~\tau_i\geq \tau_{i+1}\geq 0, \label{tau}\\ |\omega\rangle & = &
\sum_{i=1}^n \sqrt{\omega_i} |i_A~i_B\rangle, ~~~\omega_i\geq
\omega_{i+1}\geq 0,
\end{eqnarray}
and let us consider the overlap or fidelity $F_{U,V} \equiv |\langle
\tau|\omega_{U,V}\rangle|^2$ between $\tau$ and a third vector
$\omega_{U,V} \equiv (U \otimes V) \omega$, where $U$ and $V$ are any
two local unitaries on Alice's and Bob's subsystems,
respectively. Then
\begin{eqnarray}
\max_{U\otimes V} F_{U,V} = \left(\sum_{i=1}^n \sqrt{\tau_i\omega_i}\right)^2,
\label{maxover}
\end{eqnarray}
the maximal overlap corresponding precisely to the case $U=V= I$,
$\omega_{U,V}=\omega$.

%
%

{\bf Proof:} Let us begin by re-expressing $\tau ,\omega $ in the form
\cite{Jozsa94c}
\begin{equation}
\left| \tau \right\rangle = I \otimes \sigma ^{\tau }\left| \alpha
\right\rangle ;\left| \omega \right\rangle = I \otimes \sigma ^{\omega
}\left| \alpha \right\rangle ,
\end{equation}
where $\sigma ^{\tau },\sigma ^{\omega }$ are the diagonal $n\times n$
matrices constructed from the ordered Schmidt coefficients of $\tau
,\omega $
(i.e., $\sigma _{ii}^{\tau }=\sqrt{\tau _{i}}$) and $\alpha
=\sum_{i=1}^{n}\left| i_{A}i_{B}\right\rangle $ is the unnormalized
maximally entangled state. The overlap between $\tau $ and any vector $%
\omega_{U,V}$ obtained from $\omega $ by local unitary rotations is
then        \begin{eqnarray}
\left| \left\langle \tau \right| U\otimes V\left| \omega \right\rangle
\right| ^{2} &=&\left| \left\langle \alpha \right| \left( U \otimes %
\sigma ^{\tau }V\sigma ^{\omega }\right) \left| \alpha \right\rangle \right|
^{2}  \nonumber \\
&=&\left| \left\langle \alpha \right| \left( I \otimes \sigma ^{\tau
}V\sigma ^{\omega }U^{T}\right) \left| \alpha \right\rangle \right| ^{2}
\nonumber \\
&=&\left| \mbox{Tr}\left( \sigma ^{\tau }V\sigma ^{\omega }U^{T}\right) \right| ^{2}
\label{unitover}
\end{eqnarray}
where we have used the easily verified observations that $U \otimes I
\left| \alpha \right\rangle = I \otimes U^{T}\left| \alpha
\right\rangle $ and $\left\langle \alpha \right| \left( I \otimes
  A\right) |\alpha\rangle = \mbox{Tr}[A]$. The desired result follows
directly from Problem III.6.12 in \cite{Bhatia97a}.  Alternatively, a
sketch of the remainder of the proof is as follows.  First, rewrite
\begin{equation}
\left| \mbox{Tr}\left( \sigma ^{\tau }V\sigma ^{\omega }U^{T}\right) \right|
=\left| \mbox{Tr}\left( \sqrt{\sigma ^{\tau }}V\sqrt{\sigma ^{\omega }}\sqrt{\sigma
^{\omega }}U^{T}\sqrt{\sigma ^{\tau }}\right) \right| .
\end{equation}
By the Cauchy-Schwarz inequality $\left|\mbox{Tr}(A^{\dag }B)\right|
\leq \sqrt{ \mbox{Tr}\left( A^{\dag }A\right) \mbox{Tr}\left( B^{\dag
}B\right) }$, we have then
\begin{equation}
\left|\mbox{Tr}\left( \sigma ^{\tau }V\sigma ^{\omega }U^{T}\right) \right| \leq
\sqrt{\mbox{Tr}\left( \sigma ^{\omega }V^{\dag }\sigma ^{\tau }V\right) \mbox{Tr}\left(
\sigma ^{\tau }U^{*}\sigma ^{\omega }U^{T}\right) }.  \label{CSchwartz}
\end{equation}
Define $C \equiv V^{\dag }\sigma ^{\tau }V$. Since $\sigma ^{\omega }$
is diagonal, we have $\mbox{Tr}\left( \sigma ^{\omega }C\right)
=\mbox{Tr}\left( \sigma ^{\omega }\mbox{diag}(C)\right) $, where
$\mbox{diag}(C)$ is obtained by retaining only the diagonal elements
of $C$. Now, since $\sigma ^{\tau }$ diagonalizes $C$, Schur's Theorem
(\cite{Marshall79a}, Theorem 9.B.1) implies that there exist
permutation operators $P_{i}$ such that
\begin{equation}
\mbox{diag}(C)=\sum_{i}p_{i}P_{i}\sigma ^{\tau }P_{i}^{\dag }.
\end{equation}
It follows that
\begin{eqnarray}
\mbox{Tr}\left( \sigma ^{\omega }C\right)  &=&\sum_{i}p_{i}\mbox{Tr}\left( \sigma ^{\omega
}P_{i}\sigma ^{\tau }P_{i}^{\dag }\right)    \nonumber \\
&\leq &\sum_{i}p_{i}\mbox{Tr}\left( \sigma ^{\omega }\sigma ^{\tau }\right)
=\mbox{Tr}\left( \sigma ^{\omega }\sigma ^{\tau }\right) .
\end{eqnarray}
where the inequality follows from the observation that $x_{1}\leq
x_{2}$ and $y_{1}\leq y_{2}$ imply that $x_{1}y_{2}+x_{2}y_{1}\leq
x_{1}y_{1}+x_{2}y_{1}$, and so $\mbox{Tr}(\sigma^{\omega}P_i
\sigma^{\tau} P_i^{\dagger}) \leq
\mbox{Tr}(\sigma^{\omega}\sigma^{\tau})$.  Similarly,
$\mbox{Tr}(\sigma^{\tau}U^*\sigma^{\omega}U^T) \leq
\mbox{Tr}(\sigma^{\omega} \sigma^{\tau})$.  Substituting these results
in~(\ref{CSchwartz}) and then into~(\ref{unitover}), we finally obtain
\begin{equation}
\left| \left\langle \tau \right| U\otimes V\left| \omega \right\rangle
\right| ^{2}\leq \mbox{Tr}^{2}\left( \sigma ^{\omega }\sigma ^{\tau }\right)
\end{equation}
which is precisely the overlap between $\tau $ and $\omega$ given
in equation~(\ref{maxover}) .

$\Box$

%
%

{\bf Lemma 2:} Among all possible local transformations of the
bipartite pure state $\psi$, $\psi \rightarrow \{p_k, \rho_k\}$, a
deterministic one, $\psi \rightarrow \xi$, into some pure state $\xi$
can always be found which achieves the most faithful transformation
with respect to the target state $\phi$.

%
%

{\bf Proof:} Because of the linearity of the trace Tr[.], the overlap
Tr[$|\phi\rangle \langle \phi|\rho$] between $\phi$ and a mixed state
$\rho$ equals the average overlap between $\phi$ and any ensemble
realizing $\rho$. Therefore we can consider, without loss of
generality (compare the discussion around equation~(\ref{mixing})),
just local transformations $\psi \rightarrow \{p_k, \xi_k\}$ into pure
states $\xi_k$, with squared Schmidt coefficients $\gamma^k_i
\geq\gamma^k_{i+1} \geq 0$. By Lemma~1, the average fidelity $\bar{F}$
with the target state $\phi$ of equation~(\ref{phi}) satisfies
\begin{eqnarray}
\bar{F} \leq \sum_k p_k \left(\sum_{i=1}^n
\sqrt{\gamma^k_i\beta_i}~\right)^2.
\end{eqnarray}
Moreover, it follows from equations~(\ref{iff2}) and (\ref{ensemble}) that the pure
state $\bar{\xi}$, defined as
\begin{eqnarray}
|\bar{\xi}\rangle \equiv \sum_{i=1}^n \sqrt{\sum_k p_k \gamma^k_i}
 |i_A'~i_B'\rangle ,
\end{eqnarray} with the same Schmidt basis as the target state $\phi$, can be obtained
deterministically from $\psi$ in equation~(\ref{psi}). The concavity
of Uhlmann's fidelity $F(\rho_1,\rho_2)
\equiv($Tr$\sqrt{\sqrt{\rho_1}\rho_2\sqrt{\rho_1}}~)^2$
\cite{Uhlmann76a} implies that the overlap between $\bar{\xi}$ and the
target state $\phi$ is an upper bound on $\bar{F}$,
\begin{eqnarray} 
& &
\bar{F}\leq \sum_k p_k \left(\sum_{i=1}^n\sqrt{\gamma^k_i\beta_i}\right)^2
\leq\left(\sum_{i=1}^n\sqrt{\sum_kp_k\gamma^k_i~
\beta_i}\right)^2. \nonumber \\
& & 
\label{Uhlmannconcavity}
\end{eqnarray}
%
%
More precisely, define diagonal $n\times n$ matrices $\sigma^{\phi}$,
$\sigma^{\xi_k}$ and $\sigma^{\bar{\xi}}$, constructed from the square
of the ordered Schmidt coefficients of $\phi$, $\xi_k$ and
$\bar{\xi}$, respectively (e.g.  $\sigma^{\phi}_{ii}=\beta_i$). Then
the second inequality in equation (\ref{Uhlmannconcavity}) is
equivalent to
\begin{equation}
\sum_k p_k F(\sigma^{\xi_k},\sigma^{\phi}) \leq F(\sigma^{\bar{\xi}},
\sigma^{\phi}),
\end{equation}
which corresponds to concavity of the fidelity since by construction
 $\sigma^{\bar{\xi}} = \sum_k p_k \sigma^{\xi_k}$.

$\Box$

Lemma~2 implies that we need focus only on deterministic conversions
into a final pure state $\xi$.  We assume, without loss of generality,
that $n$ (the dimension of the local Hilbert spaces) is the greatest
of the number of non-vanishing Schmidt coefficients of the initial
state $\psi$ and the target state $\phi$. We need to introduce some
notation before we finally present the most faithful local
conversion. Let us then call $l_1$ the smallest integer $\in [1,n]$
such that
\begin{eqnarray}
\frac{E_{l_1}^{\psi}}{E_{l_1}^{\phi}} = \min_{l\in[1,n]}
\frac{E_{l}^{\psi}}{E_{l}^{\phi}} \equiv ~r_1~~ (\leq 1).
\label{l1}
\end{eqnarray}
It may happen that $l_1 = r_1 = 1$. If not, it follows from the equivalence
\begin{eqnarray}
\frac{a}{b} < \frac{a+c}{b+d} \,\,\,\, \Leftrightarrow \,\,\,\,
\frac{a}{b} < \frac{c}{d}~~~~~(a,b,c,d > 0)
\end{eqnarray}
that for any integer $k \in [1, l_1-1]$
\begin{eqnarray}
\frac{E_{k}^{\psi} - E_{l_1}^{\psi}}{E_{k}^{\phi} - E_{l_1}^{\phi}} > r_1.
\end{eqnarray}
Let us then define $l_2$ as the smallest integer $\in [1,l_1-1]$ such that
\begin{eqnarray}
r_2\equiv \frac{E_{l_2}^{\psi} - E_{l_1}^{\psi}}{E_{l_2}^{\phi} -
E_{l_1}^{\phi}} = \min_{l\in[1,l_1-1]} \frac{E_{l}^{\psi} -
E_{l_1}^{\psi}}{E_{l}^{\phi} - E_{l_1}^{\phi}} ~~~ (> r_1).
\label{l2}
\end{eqnarray}
Repeating this process until $l_k=1$ for some $k$, we obtain a series
of $k+1$ integers $l_0 > l_1 > l_2 > \cdots >l_k$ ($l_0 \equiv n+1$)
and $k$ positive real numbers $0 < r_1 < r_2 < \ldots < r_k$, by means
of which we define our final state
\begin{eqnarray}
|\xi\rangle \equiv \sum_{i=1}^n \sqrt{\gamma_i} |i_A'~i_B'\rangle,
\label{xi}
\end{eqnarray}
where $|i_A'\rangle$, $|i_B'\rangle$ are the same as in
equation~(\ref{phi}), and
\begin{eqnarray}
\gamma_i \equiv r_j\beta_i ~~~\mbox{ if } i \in [l_j,
l_{j-1}-1],
\end{eqnarray}
that is,
\begin{eqnarray}
\vec{\gamma} = \left[ \begin{array}{c}
r_k\left[ \begin{array}{c}
    \beta_{l_k} \\ \vdots \\ \beta_{l_{k-1}-1}
    \end{array} \right] \\
\vdots \\
r_2\left[ \begin{array}{c}
    \beta_{l_2} \\ \vdots \\ \beta_{l_1-1}
    \end{array} \right]\\
r_1\left[ \begin{array}{c}
    \beta_{l_1} \\ \vdots \\ \beta_{l_0-1}
    \end{array} \right]
\end{array} \right].
\label{defixi}
\end{eqnarray}
By construction $\gamma_i \geq \gamma_{i+1}$ and
\begin{eqnarray}
E_l^{\psi} \geq E_l^{\xi}~~~~\forall l\in[1,n],
\end{eqnarray}
so the vector $\vec{\alpha}$ is majorized by the vector
$\vec{\gamma}$, $\vec{\alpha} \prec \vec{\gamma}$.  According to
condition~(\ref{iff}) the local strategy presented in
\cite{Nielsen99a} will indeed allow the parties to obtain the state
$\xi$ from $\psi$ with certainty. Now, let us define positive
quantities
\begin{eqnarray}
A_j \equiv E_{l_j}^{\psi} - E_{l_{j-1}}^{\psi} &=& \sum_{i=l_j}^{l_{j-1}-1} \alpha_i
\label{Ai}~~~~(E_{l_0}^{\psi}\equiv 0),\\
B_j \equiv E_{l_j}^{\phi} - E_{l_{j-1}}^{\phi} &=&
\sum_{i=l_j}^{l_{j-1}-1} \beta_i \label{Bi}~~~~(E_{l_0}^{\phi}\equiv 0).
\end{eqnarray}
Then the fidelity between the final state $\xi$ and the target state
$\phi$ reads, in terms of the initial and target states,
\begin{eqnarray}
|\langle \xi | \phi\rangle|^2 = \left( \sum_{j=1}^k \sqrt{A_j B_j}
\right)^2.
\label{optfid}
\end{eqnarray}

Without loss of generality, Lemma~1 allows us to assume that any other
possible final state $\xi'$ has the same Schmidt basis as the target
state $\phi$ and squared Schmidt coefficients $\gamma_i' \geq
\gamma_{i+1}' \geq 0$, so by the Cauchy-Schwarz inequality
$(\sqrt{x_1y_1}+\sqrt{x_2y_2})^2 \leq (x_1+x_2)(y_1+y_2)$,
$x_1,x_2,y_1,y_2 \geq 0$),
\begin{eqnarray}
F_{\xi'} & \equiv & F(\xi',\phi) \equiv \left(\sum_{i=1}^n \sqrt{
\gamma_i'\beta_i}\right)^2 \\ & \leq & \left(\sum_{j=1}^k \left(
\sum_{i=l_j}^{l_{j-1}-1}\gamma_i' \right)^{1/2} \sqrt{B_j}\right)^2,
\label{ineq}
\end{eqnarray}
where $l_j$ ($j=1, \cdots, k$) have been defined in
equations~(\ref{l1})-(\ref{l2}). Now, recall that
$\sum_{i=l_j}^{l_{j-1}-1}\gamma_i' =
E^{\xi'}_{l_j}-E^{\xi'}_{l_{j-1}}$, and that the local and
deterministic character of the conversion $\psi \rightarrow \xi'$
implies that $E^{\psi}_l \geq E^{\xi'}_l~$ ($l=1,\cdots,n$). We can
therefore define $a_{j}$ as
\begin{eqnarray}
a_{j} \equiv E^{\psi}_{l_j} - E^{\xi'}_{l_j};~~a_0 \equiv 0.
\end{eqnarray}
The condition $\alpha \prec \gamma'$ implies that $a_j \geq 0$ for
each $j$.  We may rewrite equation~(\ref{ineq}) in terms of the $a_j$
and the $A_j$ introduced in equation~(\ref{Ai}) as
\begin{eqnarray}
F_{\xi'} \leq \left( \sum_{j=1}^k \sqrt{A_j - a_j + a_{j-1}}
\sqrt{B_j}\right)^2 \equiv f(\vec{a}).
\end{eqnarray}
Our interest is in the behaviour of $f(\vec a)$ as a function of $\vec
a$.  We will show that in the allowed parameter region $f(\vec{a})$ is
maximized when $\vec a = 0$.  A direct computation shows that the
(tridiagonal) matrix of second derivatives of $f(\vec a)$, $(M_n)_{ij}
\equiv \frac{\partial^2 f}{\partial a_i\partial a_j}$, is negative
definite in the region ${\cal A}\subset{\cal R}^n$ defined by the
constraints $a_j \geq 0$ and $A_j - a_j + a_{j-1}\geq 0$, which
contains all relevant situations compatible with $\gamma_i'\geq
\gamma_{i+1}' \geq 0$.  Next, note that
\begin{eqnarray}
  \frac{\partial f(\vec{a})}{\partial a_j}\left|_{\vec{a}=\vec{0}}=
    \sqrt{f(\vec{0})}\left(\sqrt{\frac{B_{j+1}}{A_{j+1}}}-\sqrt{\frac{B_j}{A_j}}\right).
  \right.
\end{eqnarray}
By construction $\frac{A_j}{B_j}<\frac{A_{j+1}}{B_{j+1}}$ (compare
equations~(\ref{l1})-(\ref{l2}) and (\ref{Ai})-(\ref{Bi})), so
\begin{eqnarray}
  \frac{\partial f(\vec{a})}{\partial a_j}\left|_{\vec{a}=\vec{0}} < 0 \right.
\end{eqnarray}
It follows that the maximum of $f(\vec{a}\in {\cal A})$ occurs at
$\vec{a}=\vec{0}$, that is, when the final state $\xi'$ is precisely
the state $\xi$ as defined in equations~(\ref{xi})-(\ref{defixi}).
Therefore, we can conclude:

{\bf Theorem 3:} The maximal fidelity $F_{opt}$ achievable in a
faithful local transformation of the initial pure state $\psi$ into
the target pure state $\phi$ is given by equation~(\ref{optfid}),
\begin{eqnarray} \label{eqtn:opt_fid_2}
F_{opt} = \left(\sum_{j=1}^k \sqrt{A_jB_j}\right)^2.
\end{eqnarray}
The most faithful protocol consists in a deterministic conversion of
$\psi$ into the pure state $\xi$ as defined in
equations~(\ref{xi})-(\ref{defixi}).

\section{Discussion and applications}
\label{sec:discussion}

The next few sections apply Theorem~3 to several problems of
entanglement transformation.  Section~\ref{subsec:concentration} finds
the most faithful protocol for performing a special type of
entanglement transformation known as {\em entanglement concentration},
in which a large number of partially entangled states are transformed
into Bell pairs. This result is then applied to determine the most
faithful teleportation protocol via any given pure quantum state.
Section~\ref{subsec:dilution} finds the most faithful protocol for
performing the reverse procedure to concentration, {\em entanglement
  dilution}. Section~\ref{subsec:comparison} compares the most
faithful transformation with the optimal conclusive transformation,
and concludes that in general they are different.
Section~\ref{subsec:robust} explains how our results can be used to
demonstrate the robustness against noise of entanglement
transformation protocols for pure states, and
Section~\ref{subsec:robust_catalysis} explains this in the special
case of entanglement catalysis.

\subsection{Concentration of entanglement and optimal teleportation fidelity}
\label{subsec:concentration}

An {\em entanglement concentration protocol} \cite{Bennett96c} is a
strategy for obtaining maximally entangled states from some partially
entangled initial (pure) state $\psi$ using only LOCC. In the original
formulation of this concept, due to Bennett, Bernstein, Popescu and
Schumacher \cite{Bennett96c}, many ($N$) copies of an $n\times
n$-dimensional state $\psi$ are available, and the goal is to obtain
the largest number of $n$-states in the asymptotic limit where
$N\rightarrow\infty$. More recently, the optimal way to conclusively
concentrate the entanglement of a {\em single copy} of $\psi$ has also
been obtained \cite{Lo97b,Jonathan99a,Hardy99a}.

In this section we solve the same problem from the point of view of
faithful conversions. In this case, the goal is to determine the local
strategy that maximizes the fidelity between the single copy of $\psi$
and the maximally entangled ``$n$-state" $\psi_n$. It turns out that
the optimal strategy in this case is essentially to do nothing at
all. The only requirement is to apply the local unitary rotations that
align the Schmidt components of $\psi$ to those of $\psi_n$, in the
manner implied by Lemma~1. This result can be shown using
equations~(\ref{l1})-(\ref{defixi}).  However, a simpler derivation
can be obtained from the following argument.  First, for any pure
state $\psi$ with Schmidt coefficients $\sqrt{\alpha_1}\geq...\geq
\sqrt{\alpha_n}$, consider the function
\begin{eqnarray}
F_{max}(\psi)=\frac{1}{n}\left(\sum_{i=1}^{n}\sqrt{\alpha_i}\right)^2.
\label{eqtn:opt_concfid}
\end{eqnarray}
As has been pointed out by M.~Horodecki~\cite{Horodecki99d}, $F_{max}$
is a unitarily invariant, concave function of the reduced density
matrix $\rho_A = \mbox{Tr}_B |\psi\rangle \langle \psi|$. Following
Theorem 2 in \cite{Vidal98a}, it is therefore an entanglement monotone
for pure states. In fact, Lemma~1 shows that $F_{max} (\psi)$ is the
greatest fidelity with respect to $\psi_n$ that is achievable from
$\psi$ by local unitary rotations. Now, following Lemma 2, let $\xi$
be the most faithful approximation of $\psi_n$ obtainable from $\psi$
by LOCC. By definition then, $F_{max}(\psi)\leq F_{max}(\xi)$. On the
other hand, since $F_{max}$ is an entanglement monotone, we must also
have $F_{max}(\psi)\geq F_{max}(\xi)$. These quantities are therefore
equal, which implies that the optimally faithful strategy can be
achieved using only local unitary rotations.

It is interesting to note that $F_{max}$ can also have another
interpretation. It is equivalent to the {\em robustness of
entanglement} $R(\psi)$, an entanglement monotone that was
comprehensively studied in \cite{Vidal98a}. $R(\psi)$ is defined as
the minimal amount of separable noise that has to be mixed with
state $\psi$ in order to wash out its quantum correlations
completely. For pure states, its value reads
$R(\psi)=nF_{max}(\psi)-1$.

An important consequence of determining $F_{max}$ is that it also
allows us to determine the optimal {\em fidelity of teleportation}
via $\psi$. Recall that perfect teleportation of an unknown
$n$-dimensional state can be realized only if an ``$n$-state'' is
shared between Alice and Bob \cite{Bennett93a}. For a more general
initially shared state $\psi$, one must admit some imperfection in
the procedure. As with entanglement transformations, it is possible
to consider two approaches to imperfect teleportation: on the one
hand, conclusive teleportation strategies seek to maximize the
likelihood of achieving ideal teleportation, but also allow for the
possibility of failure \cite{Mor96a}. On the other hand, faithful
strategies seek to maximize the so-called fidelity of
teleportation. For any given teleportation strategy ${\cal T}$,
this quantity is naturally defined \cite{Horodecki99c} as the
average overlap between Alice's initial state $\phi$ and the final
teleported state obtained by Bob
\begin{eqnarray}
f(\Lambda_{{\cal T},\psi})= \int d\phi \langle \phi| \Lambda_{{\cal
T},\psi}(|\phi\rangle \langle \phi|) |\phi\rangle,
\end{eqnarray}
where $\Lambda_{{\cal T},\psi}$ is the trace-preserving quantum
operation that maps the initial state onto the teleported one (a
construction for this operation may be found in \cite{Nielsen97b}).

Recently, a connection has been found between this quantity and
faithful entanglement concentration procedures \cite{Horodecki99c}.
It has been shown that, for any given initial state $\rho$ (pure or
mixed) in $n\times n$-dimensional Hilbert space, the maximum value
of $f$ over {\em all possible} teleportation protocols implemented
using LOCC is given by
\begin{eqnarray}
f_{max}(\rho) = \frac{F_{max}(\rho) n + 1}{n + 1}.
\label{eqtn:telepfid}
\end{eqnarray}
Here, $F_{max}(\rho)$ is precisely the maximum fidelity that can be
achieved between $\rho$ and an ``$n$-state'' under a trace-preserving
quantum operation implemented via local operations and classical
communication. In general, it is not yet known how to calculate this
quantity. However, in the case of a pure initial state $\rho = \psi$,
its value is the one found in equation (\ref{eqtn:opt_concfid})
above. The maximum fidelity of teleportation via $\psi$ is then also
immediately determined via equation (\ref{eqtn:telepfid}):
\begin{eqnarray}
f_{max}(\psi) = \frac{\left(\sum_{i=1}^{n}\sqrt{\alpha_i}\right)^2  + 1}{n + 1}.
\end{eqnarray}

A `most faithful' teleportation protocol that achieves this limit
has also been described in \cite{Horodecki99c}. For any initial
state $\rho$, its first step requires transforming $\rho$ into the
most faithful achievable approximation of an $n$-state. In the case
of a pure state $\psi$, we now know that this is done merely by the
Schmidt-basis alignment described above. The remainder of the
protocol requires then only a so-called `$U\otimes U^*$ twirling'
\cite{Horodecki99c} of the state (resulting in a Werner state
\cite{Werner89a}), followed by applying the standard teleportation
procedure \cite{Bennett93a}. We therefore have now an explicit
protocol for realizing optimally faithful teleportation via pure
states.

%
%
\subsection{Entanglement Dilution}
\label{subsec:dilution}

We now consider the reverse process to entanglement concentration,
entanglement dilution \cite{Bennett96c}. In this case, the parties
start out with some $m$-state $\psi_m$ and aim at obtaining a final,
less entangled state $\phi$, constituted of $N$ copies of some
smaller-dimensional state $\chi$, i.e. $\phi=\chi^{\otimes N}$. If the
number of non-vanishing Schmidt coefficients of $\chi$ is greater than
$\root{N}\of{m}$, then this exact transformation is not possible at
all --- not even with only some probability of success --- since
$\phi$ has fewer Schmidt components than $\psi$, $m < n$ \cite{Lo97b}.
In this case, it is interesting to consider the most faithful
approximation to $\chi^{\otimes N}$ that can be achieved.

Let $|\phi\rangle= \sum_{i=1}^n\sqrt{\beta_i} |i_A~i_B\rangle$. The
most faithful approximation $\xi$ to $\phi$ that can be obtained from
$\psi_m$ by LOCC is determined using
equations~(\ref{l1})-(\ref{defixi}) as follows: from
equation~(\ref{l1}), we have $r_1=0$, $l_1 = m+1$. Equation~(\ref{l2})
gives $r_2=\left(m\sum_{i=1}^{m}\beta_i\right)^{-1}$, $l_2=1$. It
follows that
\begin{eqnarray}
|\xi\rangle = \frac{1}{\sqrt{\sum_{i=1}^m \beta_i}}
\sum_{i=1}^m\sqrt{\beta_i}|i_A~i_B\rangle,
\end{eqnarray}
and the corresponding optimal fidelity~(\ref{eqtn:opt_fid_2}) simply
reads
\begin{eqnarray}
F_{opt} = \sum_{i=1}^{m} \beta_i.
\end{eqnarray}
In other words, the best approximation to the target state $\phi$
is the state of highest norm that can be obtained by projecting
$\phi$ onto an $m\times m$-dimensional subspace.

In \cite{Bennett96c} the problem of optimal entanglement dilution was
solved in the asymptotic limit $m,N\rightarrow\infty$. In this regime,
the dilution procedure can actually be realized with 100\%
efficiency. The protocol realizing this is well-defined for any finite
values of $m,N$. It consists essentially in identifying the subspace
of $\phi$ spanned by its $m$ largest Schmidt components and then using
the $m$-state $\psi_m$ to teleport half of this over to Bob. It can be
easily verified that the resulting fidelity with respect to $\phi$ is
given precisely by the expression above. This then shows that not only
does this protocol approach fidelity 1 as $m,N\rightarrow\infty$, but
it is also optimal for any finite values of these quantities.

\subsection{Faithful versus conclusive transformations}
\label{subsec:comparison}

Suppose Alice and Bob's aim is to transform the state $\psi$ into the
state $\phi$. We have found the optimal fidelity with which this
transformation can be accomplished. A natural question to ask is how
this faithful conversion strategy compares with the optimal conclusive
strategy --- the one that maximizes the probability of successful
conversion \cite{Vidal99a}. A first observation is that the latter is
in general {\em not} also the most faithful strategy. This follows
since the optimal conclusive strategy will not usually succeed with
100\% probability, whereas Lemma~2 shows that the fidelity with
respect to $\phi$ is always maximized by means of a deterministic
transformation. A simple example is the case of a 2-qubit system
initially in a partially entangled state $a|00\rangle+b|11\rangle$,
with $a>b>0$. As we have seen above, the most faithful strategy for
converting it into the maximally entangled 2-state is simply to do
nothing, which corresponds to a fidelity of $\frac{1}{2}+ab$. On the
other hand, the optimal conclusive transformation, which succeeds with
probability $2b^2$ \cite{Lo97b,Vidal99a}, results in an average
fidelity of $\frac{1}{2}+b^2$, which is strictly less than was
achieved by the most faithful transformation.

We also note the surprising fact that, in all cases, realizing the
most faithful conversion does not diminish in any way Alice and Bob's
chances of conclusively obtaining the target state. This follows since
the final state $\xi$ in equations~(\ref{xi})-(\ref{defixi}) is
precisely the same as the intermediate state $\Omega$ in the optimal
conclusive protocol presented in \cite{Vidal99a}, equations
(10)-(14). This means then that no probability of success is lost
during a most faithful conversion, that is,
\begin{equation}
P(\psi\rightarrow\phi) = P(\xi\rightarrow\phi).
\end{equation}
In other words, the parties may postpone their decision on whether or
not they wish to risk their initial state in a conclusive
transformation into $\phi$, while obtaining already the most faithful
approximation to $\phi$.

%
%
\subsection{Robustness of transformations}
\label{subsec:robust}

Up to this point, our discussion has assumed that the initial state
$\psi$ shared by Alice and Bob is pure. Suppose, however, that $\psi$
is corrupted a little before it is made available to Alice and Bob, so
they receive a density matrix $\rho$ instead.  What can we say about
the possibility of transforming $\rho$ into a target state $\phi$?
This section establishes upper and lower bounds on the fidelity with
which the transformation $\rho \rightarrow \phi$ may be accomplished,
and the next section explains how these results may be used to analyze the
robustness of effects like entanglement catalysis\cite{Jonathan99c}.

%
%
Our results are most easily presented using the {\em trace
distance}, a metric on Hermitian operators defined by $T(A,B)
\equiv \mbox{Tr}(|A-B|)$, where $|X|$ denotes the positive square root of
the Hermitian matrix $X^2$.  Ruskai \cite{Ruskai94a} has shown that
the trace distance contracts under physical processes.  More
precisely, if $\rho$ and $\sigma$ are any two density operators,
and if $\rho' \equiv {\cal E}(\rho)$ and $\sigma' \equiv {\cal
E}(\sigma)$ denote states after some physical process represented
by the (trace-preserving) quantum operation ${\cal E}$ occurs, then
\begin{eqnarray} \label{eqtn:contractivity}
T(\rho',\sigma') \leq T(\rho,\sigma).
\end{eqnarray}
We will use $T(\psi,\phi)$ to denote the trace distance between the
density matrices $|\psi\rangle\langle \psi|$ and $|\phi\rangle\langle
\phi|$.  For pure states the trace distance and the fidelity are
related by a simple formula,
\begin{eqnarray} \label{eqtn:trace_fidelity}
T(\psi,\phi) = 2 \sqrt{1-F(\psi,\phi)}.
\end{eqnarray}

%
%
Returning to the problem of entanglement transformation, suppose
$\psi$ is a pure state that we wish to transform into a pure state
$\phi$.  Let $T(\psi \rightarrow \phi)$ denote the minimal trace
distance that can be achieved by such a transformation; this is easily
found by substituting~(\ref{eqtn:opt_fid_2})
into~(\ref{eqtn:trace_fidelity}).  We will provide upper and lower
bounds on $T(\rho \rightarrow \phi)$, the minimal trace distance to
$\phi$ that may be achieved by a protocol starting with the state
$\rho$, and using local operations and classical communication.

%
%
Suppose we start with the state $\rho$, and apply the protocol that
most faithfully transforms $\psi$ into $\phi$.  Define $\rho'$ to be
the result of applying this protocol to $\rho$, and $\psi'$ the result
of applying the protocol to $\psi$.  Then since this is just one
possible protocol, not necessarily optimal, for transforming $\rho$
into $\phi$, we must have
\begin{eqnarray}
T(\rho \rightarrow \phi) \leq T(\rho',\phi).
\end{eqnarray}
By the metric property of the trace distance,
\begin{eqnarray} \label{eqtn:metric}
T(\rho',\phi) \leq T(\rho',\psi') + T(\psi',\phi).
\end{eqnarray}
But by the contractivity property~(\ref{eqtn:contractivity}) we have
$T(\rho',\psi') \leq T(\rho,\psi)$, and the choice of protocol ensures
that $T(\psi',\phi) = T(\psi \rightarrow \phi)$.
Thus~(\ref{eqtn:metric}) implies
\begin{eqnarray}
T(\rho \rightarrow \phi) \leq T(\rho,\psi)+T(\psi \rightarrow \phi),
\end{eqnarray}
which is an upper bound on $T(\rho \rightarrow \phi)$ in terms of the
easily calculated quantities $T(\rho,\psi)$ and $T(\psi \rightarrow
\phi)$.

%
%
A lower bound on $T(\rho \rightarrow \phi)$ may be obtained by a
similar technique.  Suppose $\rho''$ and $\psi''$ are the states
obtained from $\rho$ and $\psi$, respectively, by applying the optimal
transformation protocol for obtaining $\phi$ from $\rho$.  Then we
must have
\begin{eqnarray}
T(\psi \rightarrow \phi) \leq T(\psi'',\phi).
\end{eqnarray}
By the metric property, $T(\psi'',\phi) \leq
T(\psi'',\rho'')+T(\rho'',\phi)$.  By contractivity, $T(\psi'',\rho'')
\leq T(\psi,\rho)$, and by the choice of protocol, $T(\rho'',\phi) =
T(\rho \rightarrow \phi)$.  Thus
\begin{eqnarray}
T(\psi \rightarrow \phi) \leq T(\rho,\psi)+T(\rho \rightarrow \phi),
\end{eqnarray}
which provides a lower bound on $T(\rho \rightarrow \phi)$.  Combining
upper and lower bounds on $T(\rho \rightarrow \phi)$ into a single
equation we have the useful inequality
\begin{eqnarray} \label{eqtn:mixed_bounds}
\left|T(\rho \rightarrow \phi)- T(\psi \rightarrow \phi)\right| \leq T(\rho,\psi).
\end{eqnarray}
We note in passing that the same method may be used to prove that for
any quadruple of quantum states $\rho_1,\rho_2,\sigma_1,\sigma_2$ the
following more general inequality holds,
\begin{eqnarray} \label{eqtn:mixed_bounds_2}
  |T(\rho_1 \rightarrow \sigma_1)-T(\rho_2 \rightarrow \sigma_2)| \leq
  T(\rho_1,\rho_2)+T(\sigma_1,\sigma_2).
\end{eqnarray}
This inequality is of especial use in the case where, for example,
$\rho_2$ and $\sigma_2$ are pure states, since then Theorem~3 allows
$T(\rho_2 \rightarrow \sigma_2)$ to be calculated explicitly,
and~(\ref{eqtn:mixed_bounds_2}) then bounds the quantity
$T(\rho_1\rightarrow \sigma_1)$, which we do not know how to calculate
exactly in general.

\subsection{Example: robustness of entanglement catalysis}
\label{subsec:robust_catalysis}

%
%
As an illustration of the usefulness of the
inequality~(\ref{eqtn:mixed_bounds}), we study the robustness of
the phenomenon of {\it entanglement catalysis} \cite{Jonathan99c}
under the presence of initial noise. First let us recall the nature
of this effect: it is sometimes the case that, although Alice and
Bob cannot deterministically transform $\psi$ into $\phi$ by local
operations and classical communication, there exist {\em catalyst}
entangled states $\eta$ such that $\psi\otimes\eta$ can be
transformed into $\phi\otimes\eta$ by local operations and
classical communication. More generally, partial catalyst states
may exist that improve the efficiency of the conversion from $\psi$
into $\phi$, although not to $100\%$. In \cite{Jonathan99c} this
effect was studied from the point of view of conclusive
conversions: partial catalysts were seen to improve the probability
of conclusively obtaining $\phi$ from $\psi$.  Another point of
view, along the lines of the present work, is to regard them as
reducing the minimal trace distance achievable in a faithful
conversion:
\begin{eqnarray}
T(\psi\otimes\eta \rightarrow \phi\otimes\eta) < T(\psi\rightarrow\phi)
\end{eqnarray}
We can now ask whether this improvement survives in the presence of a
distortion of the states involved. Suppose for instance that the
initial state and catalyst are subject to some noise, so that instead
of $\psi\otimes\eta$ we have in fact a mixed state $\rho$ which is
merely close to $\psi \otimes \eta$.  Taking the trace distance
$\varepsilon=T(\rho,\psi\otimes\eta)$ as a measure of the magnitude of
the noise, we can then ask how small $\varepsilon$ has to be if the
catalytic effect is to be preserved.

From equation~(\ref{eqtn:mixed_bounds}) we have
\begin{eqnarray}
& & T(\rho \rightarrow \phi\otimes\eta) -T(\psi\otimes\eta \rightarrow
\phi\otimes\eta) \leq T(\rho,\psi\otimes\eta)= \varepsilon. \nonumber
\\
& &
\label{errcatal}
\end{eqnarray}
Now let $\Delta T_{\eta}
= T(\psi\rightarrow\phi)- T(\psi\otimes\eta \rightarrow \phi\otimes\eta)$ be the
reduction in the trace distance achievable using the catalyst
$\eta$ when there is no initial error.  Then as long as
\begin{equation}
\varepsilon < \Delta T_{\eta},
\end{equation}
we still obtain $T(\rho\rightarrow\phi\otimes\eta) <
T(\psi\rightarrow\phi)$, and therefore a catalytic enhancement of
the fidelity obtainable via LOCC is still present.

\section{Non-local distance measures}
\label{sec:non_local}

%
%
We can use the optimality result of Theorem~3 to define notions of
fidelity and distance on the space of quantum states that measures how
different the ``non-local'' properties of those states are.  For
example, we define the {\em non-local fidelity} between pure states
$|\psi\rangle$ and $|\phi\rangle$ by
\begin{eqnarray}
F_{\mbox{nl}}(\psi,\phi) \equiv \min(F(\psi \rightarrow \phi),F(\phi
\rightarrow \psi)),
\end{eqnarray}
where $F(\psi \rightarrow \phi)$ is the optimal fidelity for
transforming $\psi$ to $\phi$ by LOCC, and $F(\phi \rightarrow \psi)$
is the optimal fidelity, in general different, for transforming $\phi$
into $\psi$ by LOCC.  The non-local fidelity quantifies the similarity
in quantum correlations present in $\psi$ and $\phi$.  The non-local
fidelity can be turned into a metric by using the {\em trace
  distance}.  Recall that the trace distance between density matrices
$\rho$ and $\sigma$ is defined by $T(\rho,\sigma) \equiv
\mbox{Tr}|\rho-\sigma|$.  For pure states $\psi$ and $\phi$ the trace
distance is related to the fidelity by the
formula~(\ref{eqtn:trace_fidelity}), which we reproduce here for
convenience:
\begin{eqnarray}
T(\psi,\phi) = 2\sqrt{1-F(\psi,\phi)}.
\end{eqnarray}
Analogous to the non-local fidelity we may define the non-local trace
distance,
\begin{eqnarray}
T_{\mbox{nl}}(\psi,\phi) \equiv 2
\sqrt{1-F_{\mbox{nl}}(\psi,\phi)}.
\end{eqnarray}
This is a metric on the space of pure states of a bipartite system,
where we agree to identify two states if they have the same Schmidt
coefficients.  To see the metric property, note that the non-local
distance is manifestly symmetric, and that
$T_{\mbox{nl}}(\psi,\phi) = 0$ if and only if $F(\psi \rightarrow
\phi) = 1$ and $F(\phi
\rightarrow
\psi) = 1$, which we know is true if and only if $\psi$ and $\phi$
have the same Schmidt coefficients.  All that remains is to prove the
triangle inequality,
\begin{eqnarray}
T_{\mbox{nl}}(\psi_1,\psi_3) \leq
T_{\mbox{nl}}(\psi_1,\psi_2)+T_{\mbox{nl}}(\psi_2,\psi_3)
\label{eqtn:triang}
\end{eqnarray}
To prove this, we use a construction illustrated in
Figures~\ref{fig:metric_1} and~\ref{fig:metric_2}.  Without loss of
generality, we suppose that
\begin{eqnarray}
T_{\mbox{nl}}(\psi_1,\psi_3) = T(\psi_3,\phi),
\end{eqnarray}
where $\phi$ is the best possible approximation to $\psi_3$ that may
be obtained from $\psi_1$ by local operations and classical
communication.  Furthermore, let $\phi_2$ be the best approximation to
$\psi_2$ that can be obtained from $\psi_1$ by local operations and
classical communication, and let $\phi_3$ be the best approximation to
$\psi_3$ that can be obtained from $\psi_2$ by local operations and
classical communication.  Then
\begin{eqnarray} \label{eqtn:thunderbolts}
T(\psi_2,\phi_2) \leq T_{\mbox{nl}}(\psi_1,\psi_2); \,\,\,\,
T(\psi_3,\phi_3)
\leq T_{\mbox{nl}}(\psi_2,\psi_3).
\end{eqnarray}
Furthermore, let $p_i,\phi_i'$ be the ensemble of states that results
when the protocol used to transform $\psi_2$ into $\phi_3$ is applied
to $\phi_2$ instead.  Define $\rho \equiv \sum_i p_i |\phi_i'\rangle \langle
\phi_i'|$.  Then since $\rho$ may be obtained from $\psi_1$ by local
operations and classical communication we have
\begin{eqnarray}
T_{\mbox{nl}}(\psi_1,\psi_3) & \leq & T(\rho,\psi_3) \\ & \leq &
              T(\rho,\phi_3)+T(\phi_3,\psi_3), \label{eqtn:trees}
\end{eqnarray}
where we applied the metric property of the trace distance on the
second line.  We again use the result of Ruskai \cite{Ruskai94a}
stating that $T(\cdot,\cdot)$ never decreases if the same
trace-preserving quantum operation is applied to each argument, so
$T(\rho,\phi_3) \leq T(\phi_2,\psi_2)$.  Combining this observation
with~(\ref{eqtn:trees}) and then applying~(\ref{eqtn:thunderbolts})
gives
\begin{eqnarray} T_{\mbox{nl}}(\psi_1,\psi_3)
& \leq & T(\phi_2,\psi_2)+T(\phi_3,\psi_3) \\ & \leq &
T_{\mbox{nl}}(\psi_1,\psi_2)+T_{\mbox{nl}}(\psi_2,\psi_3),
\end{eqnarray}
which is the triangle inequality~(\ref{eqtn:triang}).

%
%
Analogous constructions may be carried out for the mixed state case.
Unfortunately, general conditions for transforming one mixed state to
another by local operations and classical communication are not yet
known, so we cannot evaluate the non-local distance or non-local
fidelity in this instance.  (Note however that
equation~(\ref{eqtn:mixed_bounds}) does allow one to prove bounds on
the general non-local distance.)  In the case of mixed states there
are inequivalent measures of distance available for use in the
definition of non-local distance, such as the trace distance and the
Bures distance \cite{Bures69a}.  In general, any good measure of
distance for quantum states can be used to define a good measure of
non-local distance, provided it has a contractivity property analogous
to that for the trace distance (which, for example, the Bures distance
has).

\section{Conclusion}
\label{sec:conc}

%
%
We have found the optimal approximate schemes for transforming one
pure entangled state into another using local operations and classical
communication.  These results have been used to determine the best
possible schemes for entanglement concentration and dilution, to
determine the optimal teleportation fidelity that may be achieved when
imperfect pure state entanglement is available, and to obtain bounds
on how well entanglement can be transformed in the presence of a small
amount of noise in the initial state. This in turn allows us to
estimate how robust surprising effects such as entanglement catalysis
are against such small perturbations. Furthermore, we defined a {\em
non-local fidelity} to measure the difference in the entanglement
present in two quantum states.  This quantity is not affected by local
unitary changes to the system, and can be used to define interesting
non-local metrics on the space of entangled states.  We believe that
these results shed considerable light on the ongoing effort to develop
the notion of entanglement as a physical resource that can be employed
in a wide variety of information processing tasks.  In particular, an
understanding of approximation is crucial to the analysis of proposals
for tasks of practical interest, like the cryptographic protocol
recently proposed by Barnum \cite{Barnum99a}, whose security depends
upon the difficulty of performing certain entanglement
transformations.

\section*{Acknowledgments}

G.V. acknowledges a CIRIT grant 1997FI-00068 PG and financial support
from CIRYT contract AEN98-0431, CIRIT contract
1998SGR-00026. D.J. thanks the Brazilian agency Conselho Nacional de
Desenvolvimento Cient\'{\i}fico e Tecnol\'ogico (CNPq), the Leverhulme
Trust, EPSRC, the ORS Award Scheme and the European Union. M.A.N. was
supported by a Tolman Fellowship and by DARPA through the Quantum
Information and Computing Institute (QUIC) administered through the
ARO.

Part of this work was completed during the `Complexity, Computation
and the Physics of Information' workshop of the Isaac Newton
Institute and the ESF-QIT meeting in Cambridge, July 1999. The
authors thank the Institute and the European Science Foundation for
support during this period.

\begin{figure}
\begin{center}
\unitlength 1cm
\begin{picture}(8,3)(0,0)
\put(2,2.5){\makebox(0,0){$\psi_1$}}
\put(2.2,2.3){\vector(2,-1){3}}
\put(5.5,0.6){\makebox(0,0){$\phi$}}
\put(5.5,2.5){\makebox(0,0){$\psi_3$}}
\end{picture}
\end{center}
\caption{$\phi$ is the best approximation to $\psi_3$ that may be
obtained from $\psi_1$ by local operations and classical
communication.
\label{fig:metric_1}}
\end{figure}

\begin{figure}
\begin{center}
\unitlength 1cm
\begin{picture}(7,3.5)(0,0)
\put(1,3){\makebox(0,0){$\psi_1$}}
\put(1.2,2.8){\vector(2,-1){2}}
\put(3.5,1.7){\makebox(0,0){$\phi_2$}}
\put(3.8,1.5){\vector(2,-1){2}}
\put(6.1,0.4){\makebox(0,0){$\rho$}}
\put(3.5,3){\makebox(0,0){$\psi_2$}}
\put(3.8,2.8){\vector(2,-1){2}}
\put(6.1,1.7){\makebox(0,0){$\phi_3$}}
\put(6.1,3){\makebox(0,0){$\psi_3$}}
\end{picture}
\end{center}
\caption{$\phi_2$ is the best approximation to $\psi_2$ that can be
obtained from $\psi_1$ by local operations and classical communication.
$\phi_3$ is the best approximation to $\psi_3$ that can be obtained
from $\psi_2$ by local operations and classical communication.  $\rho$
is the (possibly mixed) state that results when the protocol
converting $\psi_2$ to $\psi_3$ is applied to
$\phi_2$. \label{fig:metric_2}}
\end{figure}


\begin{thebibliography}{10}

\bibitem{Preskill98a}
J. Preskill, Proc. Roy. Soc. A: Math., Phys. and Eng. {\bf 454},  469  (1998).

\bibitem{Bennett93a}
C.~H. Bennett {\it et~al.}, Phys. Rev. Lett. {\bf 70},  1895  (1993).

\bibitem{Bennett92a}
C.~H. Bennett, G. Brassard, and N.~D. Mermin, Physical Review Letters {\bf
  68(5)},  557  (1992).

\bibitem{Bennett96b}
C.~H. Bennett {\it et~al.}, Physical Review Letters {\bf 76},  722  (1996),
  quant-ph/9511027.

\bibitem{Bennett96c}
C.~H. Bennett, H.~J. Bernstein, S. Popescu, and B. Schumacher, Phys. Rev. A
  {\bf 53},  2046  (1996), quant-ph/9511030.

\bibitem{Bennett96a}
C.~H. Bennett, D.~P. DiVincenzo, J.~A. Smolin, and W.~K. Wootters, Phys. Rev. A
  {\bf 54},  3824  (1996), quant-ph/9604024.

\bibitem{Lo97b}
H.-K. Lo and S. Popescu, quant-ph/9707038  (1997).

\bibitem{Nielsen99a}
M.~A. Nielsen, Phys.~Rev.~Lett. {\bf 83},  436  (1999).

\bibitem{Vidal99a}
G. Vidal, Phys. Rev. Lett. {\bf 83},  1046  (1999).

\bibitem{Jonathan99a}
D. Jonathan and M.~B. Plenio, Phys. Rev. Lett. {\bf 83},  1455  (1999),
  quant-ph/9903054.

\bibitem{Hardy99a}
L. Hardy, Phys. Rev. A {\bf 60},  1912  (1999).

\bibitem{Barnum99a}
H. Barnum, quant-ph/9910072  (1999).

\bibitem{Peres93a}
A. Peres, {\em Quantum Theory: Concepts and Methods} (Kluwer Academic,
  Dordrecht, 1993).

\bibitem{Vidal98a}
G. Vidal, quant-ph/9807077  (1998).

\bibitem{Bhatia97a}
R. Bhatia, {\em Matrix analysis} (Springer-Verlag, New York, 1997).

\bibitem{Kent98a}
A. Kent, Phys.~Rev.~Lett. {\bf 81},  2839  (1998), quant-ph/9805088.

\bibitem{Jonathan99c}
D. Jonathan and M.~B. Plenio, Phys. Rev. Lett. {\bf 83},  3566  (1999).

\bibitem{Kempe99a}
J. Kempe, Physical Review A {\bf 60},  910  (1999), e-print quant-ph/9902036.

\bibitem{Fuchs99a}
C.~A. Fuchs and J. van~de Graaf, IEEE Trans. Inf. Theory {\bf 45},  1216
  (1999).

\bibitem{Jozsa94c}
R. Jozsa, J. Mod. Opt. {\bf 41},  2315  (1994).

\bibitem{Marshall79a}
A.~W. Marshall and I. Olkin, {\em Inequalities: theory of majorization and its
  applications} (Academic Press, New York, 1979).

\bibitem{Uhlmann76a}
A. Uhlmann, Reports on Mathematical Physics {\bf 9},  273  (1976).

\bibitem{Horodecki99d}
M. Horodecki, private communication, 1999.

\bibitem{Mor96a}
T. Mor, quant-ph/9608005  (1996).

\bibitem{Horodecki99c}
M. Horodecki, P. Horodecki, and R. Horodecki, Phys.~Rev.~A {\bf 60},  1888
  (1999).

\bibitem{Nielsen97b}
M.~A. Nielsen and C.~M. Caves, Phys. Rev. A {\bf 55},  2547  (1997).

\bibitem{Werner89a}
R.~F. Werner, Phys. Rev. A {\bf 40},  4277  (1989).

\bibitem{Ruskai94a}
M.~B. Ruskai, Rev. Math. Phys. {\bf 6},  1147  (1994).

\bibitem{Bures69a}
D. Bures, Transactions of the American Mathematical Society {\bf 135},  199
  (1969).

\end{thebibliography}
\end{document}